\documentclass[%
 reprint,
 amsmath,amssymb,
 aps,
 prl,
 superscriptaddress,
]{revtex4-2}

\usepackage{graphicx}
\usepackage{dcolumn}
\usepackage{bm}
\usepackage{hyperref}
\usepackage{comment}
\usepackage{xcolor}
\hypersetup{hidelinks}

\newcommand{\affTohoku}{Department of Physics, Tohoku University, Sendai 980-8578, Japan}
\newcommand{\affJAEA}{Advanced Science Research Center, Japan Atomic Energy Agency, Tokai 319-1195, Japan}
\newcommand{\affKyotoSangyo}{Kyoto Sangyo University, Kyoto, Japan}
\newcommand{\affKoreaUniv}{Department of Physics, Korea University, Seoul 02841, Republic of Korea}
\newcommand{\affUTokyo}{Department of Physics, The University of Tokyo, Tokyo 113-0033, Japan}
\newcommand{\affKyotoUniv}{Department of Physics, Kyoto University, Kyoto 606-8502, Japan}
\newcommand{\affOhioUniv}{Department of Physics \& Astronomy, Ohio University, Athens, OH 45701, USA}
\newcommand{\affNaraWomen}{Nara Women's University, Nara 630-8506, Japan}
\newcommand{\affKRISS}{Korea Research Institute of Standards and Science, Daejeon 34113, Republic of Korea}
\newcommand{\affKyungpook}{Department of Physics, Kyungpook National University, Daegu 41566, Republic of Korea}
\newcommand{\affSNU}{Department of Physics and Astronomy, Seoul National University, Seoul 08826, Republic of Korea}
\newcommand{\affKEK}{Institute of Particle and Nuclear Studies, High Energy Accelerator Research Organization (KEK), Tsukuba 305-0801, Japan}
\newcommand{\affJeonbuk}{Department of Physics Education, Jeonbuk National University, Jeonju 54896, Republic of Korea}

\begin{document}

\preprint{APS/123-QED}

\title{First Measurement of the $K^-$ Escape Cross Section in the ${}^{12}{\rm C}(K^{-},p)$ Reaction} 

\author{Fumiya~Oura}
\email{ohura.fumiya@jaea.go.jp} 
\affiliation{\affJAEA}
\affiliation{\affTohoku}
\author{Yudai~Ichikawa}
\affiliation{\affTohoku}
\affiliation{\affJAEA}
\author{Junko~Yamagata-Sekihara}
\affiliation{\affKyotoSangyo}
\author{Jung~Keun~Ahn}
\author{Sung~Wook~Choi}
\affiliation{\affKoreaUniv}
\author{Manami~Fujita}
\affiliation{\affUTokyo}
\author{Takeshi~Harada}
\affiliation{\affKyotoUniv}
\author{Shoichi~Hasegawa}
\affiliation{\affJAEA}
\author{Shuhei~Hayakawa}
\affiliation{\affTohoku}
\author{Kenneth~Hicks}
\affiliation{\affOhioUniv}
\author{Satoru~Hirenzaki}
\affiliation{\affNaraWomen}
\author{Sang~Hoon~Hwang}
\affiliation{\affKRISS}
\author{Ken'ichi~Imai}
\affiliation{\affJAEA}
\affiliation{\affKyotoUniv}
\author{Yuji~Ishikawa}
\affiliation{\affTohoku}
\author{Woo~Seung~Jung}
\affiliation{\affJAEA}
\author{Shunsuke~Kajikawa}
\affiliation{\affTohoku}
\author{Kento~Kamada}
\affiliation{\affTohoku}
\author{Byung~Min~Kang}
\affiliation{\affKoreaUniv}
\author{Shin~Hyung~Kim}
\affiliation{\affKyungpook}
\author{Tomomasa~Kitaoka}
\affiliation{\affTohoku}
\author{Jaeyong~Lee}
\affiliation{\affSNU}
\author{Jong~Won~Lee}
\affiliation{\affJeonbuk}
\author{Koji~Miwa}
\affiliation{\affTohoku}
\affiliation{\affKEK}
\author{Taito~Morino}
\affiliation{\affTohoku}
\author{Tamao~Sakao}
\affiliation{\affTohoku}
\author{Hiroyuki~Sako}
\affiliation{\affJAEA}
\author{Masayoshi~Saito}
\affiliation{\affTohoku}
\author{Susumu~Sato}
\affiliation{\affJAEA}
\author{Toshiyuki~Takahashi}
\affiliation{\affKEK}
\author{Kiyoshi~Tanida}
\affiliation{\affJAEA}
\author{Hirokazu~Tamura}
\affiliation{\affTohoku}
\affiliation{\affJAEA}
\author{Mifuyu~Ukai}
\affiliation{\affKEK}
\affiliation{\affTohoku}
\author{Shunsuke~Wada}
\affiliation{\affTohoku}
\author{Takeshi~O.~Yamamoto}
\affiliation{\affJAEA}
\author{Seongbae~Yang}
\affiliation{\affKoreaUniv}
\collaboration{J-PARC E42 Collaboration}\noaffiliation

\newcommand{\TOHOKU}{1}
\newcommand{\JAEA}{2}
\newcommand{\KYOSAN}{3}
\newcommand{\KU}{4}
\newcommand{\TOKYO}{5}
\newcommand{\KYOTO}{6}
\newcommand{\OHIO}{7}
\newcommand{\KRISS}{8}
\newcommand{\KNU}{9}
\newcommand{\SNU}{10}
\newcommand{\KEK}{11} 

\date{\today}

\begin{abstract}
We investigated the $\bar{K}$-nucleus interaction through the simultaneous measurement of the inclusive $^{12}{\rm C}(K^-, p)$ and exclusive $K^-$-escape $^{12}{\rm C}(K^-, p K^-_{esc})$ reactions at $1.8$ GeV/$c$ at J-PARC. The present measurement explicitly focuses on the $K^-$ escape process for the first time, successfully accomplishing a direct experimental determination of the imaginary part of the $K^-$ optical potential. The differential cross section for the $K^-$-escape reaction was determined to be $436 \pm 6\:(\text{stat.}) \pm 44\:(\text{syst.})~\mu\text{b/sr}$. A simultaneous likelihood fit yielded real and imaginary potential strengths of $V_0 = -72\:^{+3}_{-5}\:(\text{stat.})\:^{+0}_{-8}\:(\text{syst.})~\text{MeV}$ and $W_0 = -100\:^{+7}_{-1}\:(\text{stat.})\:^{+0}_{-16}\:(\text{syst.})~\text{MeV}$ at the nuclear center, respectively. The derived $W_0$ is significantly stronger than that predicted by theoretical models based on one-nucleon processes, suggesting possible contribution of multi-nucleon involving processes.

\end{abstract}

\maketitle 


\textbf{\textit{Introduction.---}} The understanding of the strong interaction in the non-perturbative regime remains a fundamental challenge in quantum chromodynamics (QCD). In this context, the antikaon-nucleon ($\bar{K}N$) interaction in the $I=0$ channel~\cite{Martin1981} is of particular importance due to its remarkably strong attraction---nearly an order of magnitude stronger than the nuclear force~\cite{Hyodo2013}. The two-body $\bar{K}N$ interaction has been clarified through $K^-p$ scattering data~\cite{Martin1981} and precision kaonic hydrogen X-ray measurements by the KEK, DEAR, and SIDDHARTA experiments~\cite{Iwasaki1997,Bazzi2011,Beer2005}, combined with theoretical analyses using the chiral SU(3) meson-baryon effective Lagrangian~\cite{IkedaHyodoWeise2012}, establishing the $\Lambda(1405)$ resonance as a $\bar{K}N$ quasi-bound state~\cite{Jido2003}. This strong attraction has led to the prediction of exotic $\bar{K}$-nuclear bound states~\cite{AkaishiYamazaki2002}. Indeed, the J-PARC E15 experiment observed the simplest kaonic nuclear state, $K^-pp$, with a binding energy of $47$~MeV and a width of $115$~MeV below the $\bar{K}NN$ threshold~\cite{Ajimura2019,Hashimoto2015,Yamaga2020}, providing firm evidence for a $\bar{K}$-nuclear bound system. However, the behavior of $\bar{K}$ in many-body nuclear systems remains elusive. The many-body dynamics are complicated by density-dependent effects such as multi-nucleon absorption processes; yet experimental constraints for such systems remain sparse, leaving a critical gap in our understanding of antikaons embedded in the nuclear medium. 

Historically, the $\bar{K}$-nucleus interaction has been probed primarily via kaonic atom X-ray spectroscopy across a wide range of nuclear targets~\cite{Batty1997,Friedman2007}. The energy shifts and widths of kaonic atom levels encode information on the $\bar{K}$-nucleus interaction at the nuclear surface, from which an optical potential is extracted. The interaction is typically parametrized by $U_{\text{opt}}(r) = (V_0 + iW_0)\rho(r)/\rho(0)$, where $\rho(r)$ is the nuclear density distribution. Here, $V_0$ represents the depth of the attractive real potential, while $W_0$ quantifies the strength of kaon absorption at the nuclear center.

Global analyses of the extensive X-ray data have provided valuable constraints on the $\bar{K}$-nucleus interaction, establishing the strongly attractive and absorptive nature of the optical potential. However, these analyses have also revealed a persistent ambiguity regarding the depth of the real potential. Chiral unitary models based on $t\rho$ approximation ~\cite{Kaiser1995,Waas1997}, which derive the in-medium $\bar{K}N$ amplitude from coupled-channel dynamics incorporating Pauli effects, nucleon binding and Fermi motion as well as short range
$NN$ correlations, predict a ``shallow" potential ($V_0 \sim -80$ to $-50$~MeV) but underestimate the absorptive strength due to the absence of multi-nucleon processes ($W_0 \sim -40$ to $-20$~MeV). Phenomenological fits~\cite{Batty1997,Friedman2007} with density-dependent models~\cite{AkaishiYamazaki2002}, which effectively incorporate multi-nucleon absorption through their nonlinear density dependence, accommodate both deep attraction ($V_0 \sim -200$ to $-150$~MeV) and strong absorption ($W_0 \sim -100$ to $-60$~MeV). Since kaonic atoms probe exclusively the nuclear surface region, both classes of solutions reproduce the X-ray data equally well---a deeper potential simply accommodates an additional bound state while yielding nearly identical atomic spectra---leaving the ``shallow vs. deep'' ambiguity unresolved. 

Recent high-precision X-ray measurements at J-PARC
E62~\cite{Hashimoto2022} have achieved sub-eV precision, marking a remarkable technical breakthrough. Nevertheless, global analyses incorporating these new data~\cite{Yamagata-Sekihara2024} still yield two disparate classes of potential parameters of $(-90,-120)$ and $(-280,-70) $ MeV at the normal nuclear density, 0.17~$\rm{fm^{-3}}$; a global analysis in the appendix of that work, including E62 data on $^3$He and $^4$He combined with heavier kaonic atom data, favors the shallow solution in terms of $\chi^2$, yet the ambiguity cannot be definitively resolved. This confirms that the limitation comes not from experimental sensitivity but from the intrinsic surface dominance of kaonic atom spectroscopy, motivating a direct determination from nuclear reactions that probe the full nuclear volume. 

In-flight $(K^-,N)$ reactions for $p_{K^{-}}<2~\rm{GeV}/c$ offer a promising way to probe the nuclear interior. In the quasi-free elastic scattering (QFES) process, $K^-``p" \to K^-p$ on a proton bound in a nucleus, the forward proton carries most of the momentum while the recoil $K^-$ has relatively low momentum ($\lesssim 250$ MeV/$c$) and thus undergoes significant final-state interaction with the residual nucleus. Theoretical spectra are calculated using the Green's function method~\cite{Yamagata2006}, which provides a unified treatment of both the bound-state and quasi-free continuum regions using a distorted-wave impulse approximation (DWIA). By comparing the experimental spectra with these calculations, one can extract the $V_0$ and $W_0$. An early in-flight $(K^-,N)$ measurement at KEK-PS (E548)~\cite{Kishimoto2007} reported a deep potential, but their spectra suffered from a coincidence bias and could not be directly compared with theoretical inclusive spectra. The J-PARC E05 experiment~\cite{Ichikawa2020} employed this approach through the inclusive measurement of the $^{12}{\rm C}(K^-, p)$ reaction and provided constraints favoring a ``shallow'' real potential, where the best fit parameters were $(V_0, W_0) \sim (-80, -40)$ MeV, with the energy dependence of the imaginary potential incorporated through the phase space factor $f_{\mathrm{phase}}(E)$ described in Ref.~\cite{Ichikawa2020}. However, since only the forward proton was detected in this measurement, the constraint on $W_0$ remains weak. 

A direct constraint on $W_0$ is crucial. The standard approaches, such as the $t\rho$ approximation, which construct the optical potential from the elementary $\bar{K}N$ amplitude, predict $W_0 \sim -40$ MeV based primarily on the one-nucleon absorption process ($K^-N \to Y\pi$, where $Y$ is hyperon)~\cite{Kaiser1995,Waas1997}. However, multi-nucleon processes such as $K^- NN \to YN$ are expected to enhance the absorptive strength significantly. Moreover, due to the dispersion relation, the real and imaginary parts of the potential are intrinsically coupled: a sufficiently large $W_0$ can induce a repulsive energy-level shift in kaonic atoms, partially mimicking the effect of a shallower real potential.

In this work (J-PARC E42 experiment), we performed an inclusive $^{12}{\rm C}(K^-, p)$ missing-mass spectrum measurement under the same kinematic conditions as the J-PARC E05 experiment. In coincidence with the inclusive measurement, we also detected the recoil $K^-$ that survives final-state absorption using the Superconducting Hyperon Spectrometer~(SHS), a large acceptance tracking device with a uniform magnetic field surrounding the $^{12}{\rm C}$ target. After quasi-free scattering the recoil $K^-$ must traverse the residual nucleus: if absorption is strong ($|W_0|$ large), fewer $K^-$ escape to be detected. Thus the escape-to-inclusive ratio is directly sensitive to $W_0$, whereas an inclusive measurement alone (detecting only the forward proton) constrains it only weakly. The ratio between the missing-mass spectra with and without $K^-$ coincidence is uniquely determined by the potential parameters $(V_0,W_0)$, enabling us to simultaneously determine both. Since $W_0$ has an energy dependence reflecting the change in the underlying reaction processes, the present measurement provides constraints complementary to kaonic atom spectroscopy probing the interaction near $B_K \approx 0$.

In this Letter, we report on the first experimental measurement of the $K^-$ escape process in the $^{12}{\rm C}(K^-, p)$ reaction at 1.8 GeV/$c$, providing a decisive constraint on the $\bar{K}$-nucleus potential. 

\textbf{\textit{Experiment.---}} The J-PARC E42 experiment was performed at the K1.8 beamline of the Hadron Experimental Facility at J-PARC. The inclusive spectrum is obtained from the kinematics of the incident $K^-$ and the forward proton in the QFES process; the exclusive spectrum is obtained by additionally detecting the escaping $K^-$ (denoted as $K^-_{esc}$) in coincidence. 

The observable of interest is the differential cross section as a function of the $K^-$ binding energy, $-B_K$, defined by $-B_K = M_{\text{miss}} - (M_{{}^{11}\text{B}} + M_{K^-})$, where $M_{\text{miss}}$ is the missing mass calculated from the four-momentum of the incident $K^-$ and the forward proton at an opening angle between them of $\theta_{Kp} = 4^\circ$ ($3.5^\circ < \theta_{Kp} < 4.5^\circ$).

The experimental setup consisted of three spectrometer systems: the K1.8 spectrometer (QQDQQ configuration) for the incident $K^-$ beam~\cite{Takahashi2012}; the KURAMA spectrometer---a magnetic spectrometer placed downstream of the target consisting of a large-gap dipole magnet, tracking chambers, and time-of-flight counters---for detecting forward high-momentum protons~\cite{Jung2025}; and the Superconducting Hyperon Spectrometer (SHS) surrounding the target for detecting low-momentum recoil $K^-$~\cite{Jung2025}. A $K^-$ beam with a momentum of $1.8$ GeV/$c$ was used to induce the reaction on a diamond target with a thickness of 20 mm (6.51 g/cm$^2$). For the inclusive measurement, the momentum of each incident $K^-$ was reconstructed through the K1.8 beamline spectrometer~\cite{Takahashi2012, Agari_2012} using the matrix method, and the forward proton was detected by the KURAMA spectrometer.

For the exclusive measurement, the SHS was employed to detect the $K^-_{esc}$. It consisted of a superconducting Helmholtz coil magnet (1~T)~\cite{Ahn_2023}, the Hyperon Time Projection Chamber (HypTPC)~\cite{Kim_2019}, and the Hyperon Time-of-Flight (HTOF) hodoscope. The HypTPC, located at the center of the Helmholtz coils surrounding the target, provided large angular coverage and enabled three-dimensional tracking of the low-momentum $K^-_{esc}$; together with the HTOF, it provided momentum measurement and particle identification. 

\textbf{\textit{Analysis.---}} The incident $K^{-}$ was identified using the beam aerogel \v{C}erenkov counter together with the beam time-of-flight system. The residual $\pi^{-}$ contamination in the selected beam-$K^{-}$ sample was below 0.01\%. The beam momentum reconstruction efficiency was 96.8\%. The forward proton was identified based on the mass squared calculated from the momentum, path length, and time-of-flight between the reaction vertex and the TOF wall, achieving a tracking efficiency of approximately 94\%. Details of these analyses are described in Ref.~\cite{Jung2025}. The forward proton momentum was determined by the Runge-Kutta method, providing a momentum resolution of approximately $2.5\%$ (FWHM) for $2.0$ GeV/$c$ protons. From these reconstructed momenta, the missing mass spectrum of the $^{12}{\rm C}(K^-, p)$ reaction was derived, corresponding to the binding energy $-B_K$ of the $K^-$ in the nucleus. 

Charged particle tracks of the recoil $K^-_{esc}$ in the HypTPC were reconstructed using a Kalman-filter-based method with a tracking efficiency of approximately 92\%~\cite{Jung2025} and a momentum resolution of approximately $3.0\%$ (FWHM) at $p_{K^-}\sim0.3$ GeV/$c$.

\begin{figure}[!h] 
\includegraphics[width=\columnwidth]{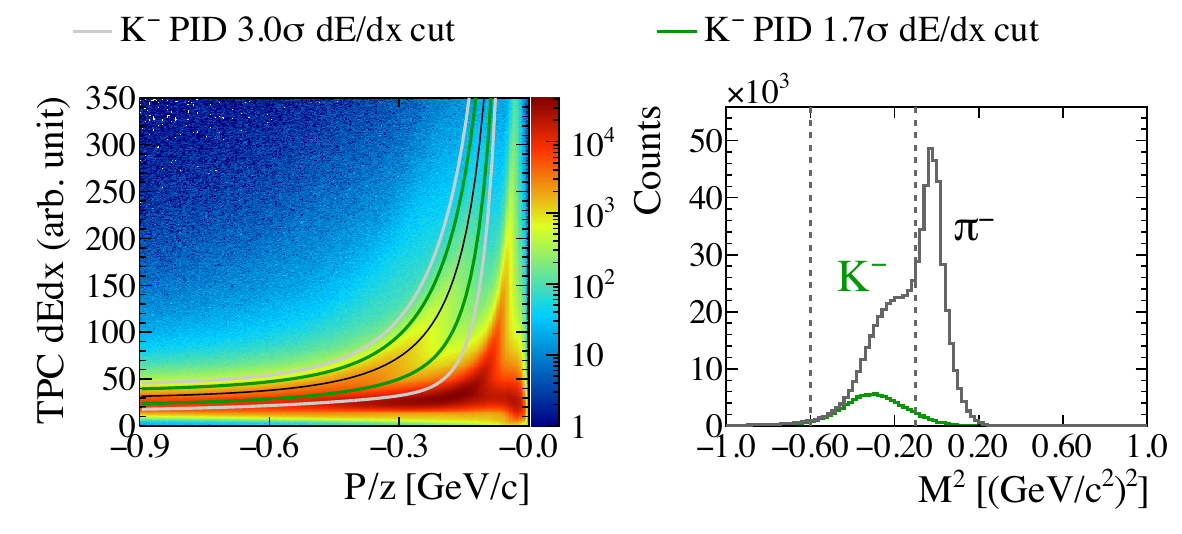}
\caption{
Particle identification for recoil $K^-_{esc}$ in the HypTPC. (a) $dE/dx$ vs.\ momentum. (b) Squared-mass distributions obtained from momentum and TOF measured in the SHS and HTOF, showing the effect of the $dE/dx$ cuts.
}
\label{fig_pid}
\end{figure}

Particle identification of the recoil $K^-_{esc}$ in the SHS was performed by combining the specific energy loss ($dE/dx$) measured in the HypTPC with the squared mass ($m^2$) calculated from the momentum reconstructed in the SHS and the time-of-flight measured with the HTOF. We selected the $dE/dx$ region within $\pm1.7\sigma$ of the expected kaon value [Fig.~\ref{fig_pid}(a)], optimized on $\mathrm{CH_2}$ data to balance purity and signal retention, and applied a squared-mass window of $0.10<m^2<0.60~{\rm GeV}^2/c^4$ [Fig.~\ref{fig_pid}(b)] around the $K^-$ peak, thereby excluding the $\pi^-$ band near $\sim0.02~{\rm GeV}^2/c^4$. The SHS--HTOF mass-squared resolution is $0.36~{\rm GeV}^2/c^4$ (FWHM) for $p_{K^-}=0.1$--$0.3$~GeV/$c$. This selection achieved a high purity for $K^-_{esc}$ of $S/(S+N)\sim99\%$, evaluated from $\mathrm{CH_2}$ data by fitting the $m^2$ spectrum after the $dE/dx$ cut with Gaussian $K^-$ and $\pi^-$ components plus a constant background in the selected mass-squared window.

\begin{figure}[!h] 
\makebox[\columnwidth][c]{\includegraphics[width=1.00\columnwidth]{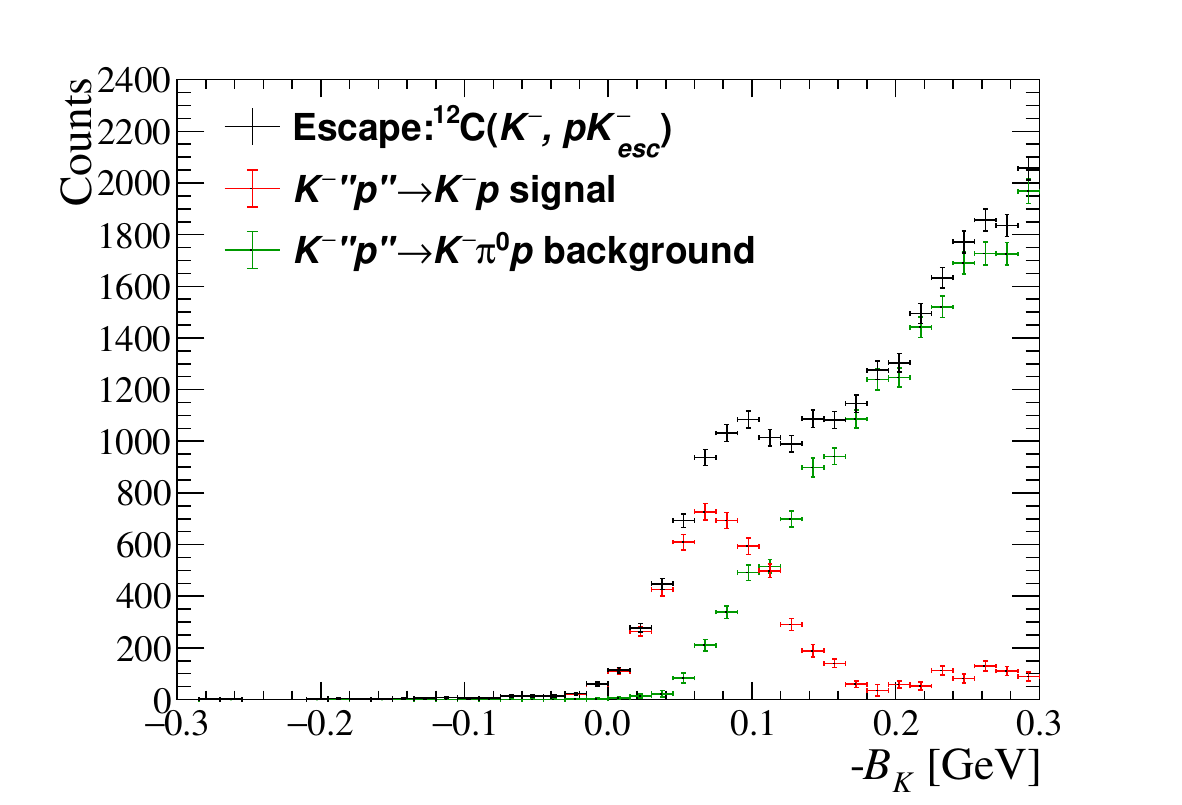}}
\caption{
$(K^-,p)$ missing-mass spectrum of $^{12}{\rm C}(K^-, pK^-_{\text{esc}})$ (black), shown in 15~MeV bins for display. Green: inelastic $K^- ``p" \to K^- \pi^0 p$ from $|\vec{P}_{\text{diff}}|$ template fits (5~MeV bins); red: background-subtracted $K^-_{esc}$ yield.
} 
\label{fig_physbg} 
\end{figure}

To obtain the exclusive escape spectrum, we selected events with a $K^-$ track in the HypTPC in coincidence with the $(K^-, p)$ reaction, i.e., $^{12}\text{C}(K^-, pK^-_{\text{esc}})$. In the $0<-B_K<0.2$~GeV region, the exclusive $K^-$ yield corresponds to $17.6\%$ of the inclusive yield and the exclusive $K^-\pi^-$ state is the only significant background with extra charged particles ($27.2\%$ relative to the exclusive $K^-$ yield). The accidental overkill ratio, defined as the fraction of genuine $K^-_{esc}$ events rejected because of an unrelated charged track in the HypTPC, is estimated to be at most $2\%$ relative to the exclusive $K^-$ yield and is incorporated in the systematic uncertainty discussed later in this section. Thus, rejecting events with any additional charged tracks strictly selects the exclusive $K^-$ channel. The $K^-$ coincidence histogram contains reactions such as quasi-free elastic scattering ($K^- ``p"\to K^-p$) and inelastic $K^- ``p"\to K^-\pi^0p$. These components were distinguished using the momentum difference $|\vec{P}_{\text{diff}}|=|\vec{P}_{\text{miss}}-\vec{P}_{K^-\text{TPC}}|$, which is shifted to larger values for the inelastic background due to the undetected $\pi^0$ momentum. The $|\vec{P}_{\text{diff}}|$ distributions of the quasi-free elastic and inelastic components were modeled by Geant4-based~\cite{Agostinelli2003} Monte Carlo simulations. For $^{12}$C, nucleon Fermi motion was implemented based on $^{12}$C$(e,e'p)$ data~\cite{Antonov2002}. We removed the inelastic background from the exclusive $K^-_{esc}$ missing-mass spectrum by fitting these simulated distributions to the data for each 5~MeV bin of $-B_K$, as shown in Fig.~\ref{fig_physbg}. In the region $0<-B_K<0.05$~GeV, the inelastic background contribution---estimated bin by bin from the $|\vec{P}_{\text{diff}}|$ template fits (green histogram in Fig.~\ref{fig_physbg})---is less than $5\%$ of the total $K^-_{esc}$ yield; the simulated $|\vec{P}_{\text{diff}}|$ distribution and measured recoil-$K^-$ momentum distribution agree with the data.

For display, the spectra in Fig.~\ref{fig_physbg} are rebinned to 15~MeV ($3\times$ the 5~MeV analysis bins); the background subtraction and optical-potential extraction use the finer binning.

We evaluated the detection efficiency and geometrical acceptance of the HypTPC and HTOF using Geant4-based~\cite{Agostinelli2003} Monte Carlo simulation incorporating electromagnetic and hadronic interactions, kaon decay, Fermi motion as above, and the PID criteria. To validate the estimate, we used a polyethylene ($\mathrm{CH_2}$) target. The detection efficiency for the recoil $K^-$ from the elementary $K^-p\to K^-p$ reaction, measured for the hydrogen component isolated by a normalized $\mathrm{CH_2}-\mathrm{C}$ subtraction with the carbon-target data scaled by the relative beam flux and number of target nuclei, was $0.22\pm0.02\,(\text{stat.})$. This agrees with the simulated value $0.24\pm0.03\,(\text{syst.})$, where $\pm0.03$ is the spread under variations of the material budget and detector-response parameters. The detection efficiency for the escaping $K^-$ in the diamond-target run was determined from simulation to be $0.125$, lower than that for the $\mathrm{CH_2}$ target ($0.242$ from simulation; quoted as $0.24\pm0.03$ in the hydrogen validation) because of the larger energy loss in the thicker diamond target (6.5~g/cm$^2$). We compared the $K^-_{esc}$ spectrum for $^{12}\rm{C}$ measured directly from the diamond target with that extracted statistically from the $\mathrm{CH_2}$ data by subtracting the hydrogen contribution determined above. After the respective efficiency corrections, the $^{12}$C $K^-_{esc}$ spectra from the diamond target and from the carbon component of the $\mathrm{CH_2}$ data agreed within $5.3\%$, confirming the target-dependent efficiency difference (0.125 vs 0.242) and justifying the simulated efficiency $0.125$ for the diamond target. The uncertainty of this efficiency is accounted for by the absolute-normalization ($8.3\%$) and target-dependent correction ($5.3\%$) systematic uncertainties. Combining these with the PID stability ($1.9\%$) and physical-background subtraction ($2.1\%$) gives a total systematic uncertainty of $10.2\%$ on the escape cross section.

The inclusive double differential cross section was obtained as $\langle d^2\sigma/(d\Omega\,dE)\rangle_{\rm inc}=N_{(K^-,p)}/(n_tN_{\rm beam}\Delta\Omega\Delta E\varepsilon_{(K^-,p)})$, where $N_{(K^-,p)}$ is the $(K^-,p)$ yield, $N_{\rm beam}$ the number of incident kaons, $n_t$ the target areal number density, $\Delta\Omega$ ($\Delta E$) the solid-angle (energy) bin width, and $\varepsilon_{(K^-,p)}$ the total detection efficiency evaluated run by run. The escape spectrum follows as $\langle d^2\sigma/(d\Omega\,dE)\rangle_{\rm esc}=R_{\rm esc}\langle d^2\sigma/(d\Omega\,dE)\rangle_{\rm inc}/\varepsilon_{K_{\rm esc}}$, where $R_{\rm esc}$ is the bin-by-bin ratio of the background-subtracted $K^-_{esc}$ yield ($N_{\rm esc}=3943\pm63$ in total) to the inclusive yield, and $\varepsilon_{K_{\rm esc}}=0.125$. Integrating over $-0.04\le-B_K\le0.2$~GeV yields $(d\sigma/d\Omega)_{\rm esc}=436\pm6\,(\text{stat.})\pm44\,(\text{syst.})~\mu\text{b/sr}$. The corresponding escape fraction---the ratio of the escape to the quasi-free elastic cross section---is $P_{\rm esc}=29\pm2\,(\mathrm{stat.})\,^{+0}_{-4}\,(\mathrm{syst.})\%$. For the previous inclusive constraint $(V_0,W_0)\sim(-80,-40)$~MeV~\cite{Ichikawa2020}, the Green's-function calculation predicts $P_{\rm esc}\sim50\%$; the measured value is substantially smaller, favoring a deeper absorptive potential.

To determine the $\bar{K}$-nucleus optical potential, we calculated inclusive and exclusive spectral templates for various combinations of $(V_0,W_0)$ within the Green's function framework~\cite{Yamagata2006}, and then performed a simultaneous fit to both datasets with $V_0$ and $W_0$ treated as free fit parameters. Prior to the fit, both spectra were corrected for their respective detection efficiencies and converted to double differential cross sections. Within the selected angular range ($3.5^\circ<\theta_{Kp}<4.5^\circ$), the efficiencies for both the inclusive and escape spectra were confirmed to be independent of $-B_K$ and therefore do not distort their shapes. For the inclusive spectrum, the fit model combines the quasi-elastic Green's function component with Geant4-based templates for inelastic backgrounds. The inelastic backgrounds include $K^-``p"$ processes constrained by J-PARC E05 results~\cite{Ichikawa2020} and $K^-``n"$ processes ($K^-``n"\to\Lambda\pi^-$, $K^-\pi^-p$, and $\Lambda\pi\pi\pi$). In these channels, protons from $\Lambda\to p\pi^-$ decay are misidentified as the forward proton of the QFES process, contaminating the $(K^-,p)$ missing-mass spectrum. The overall normalization of the inclusive spectrum was free, while the relative fractions of the inelastic components were constrained by the J-PARC E05 results~\cite{Ichikawa2020} and bubble-chamber data~\cite{VeldeWilquet1977}.

\begin{figure}[!h]
\makebox[\columnwidth][c]{\includegraphics[width=1.10\columnwidth]{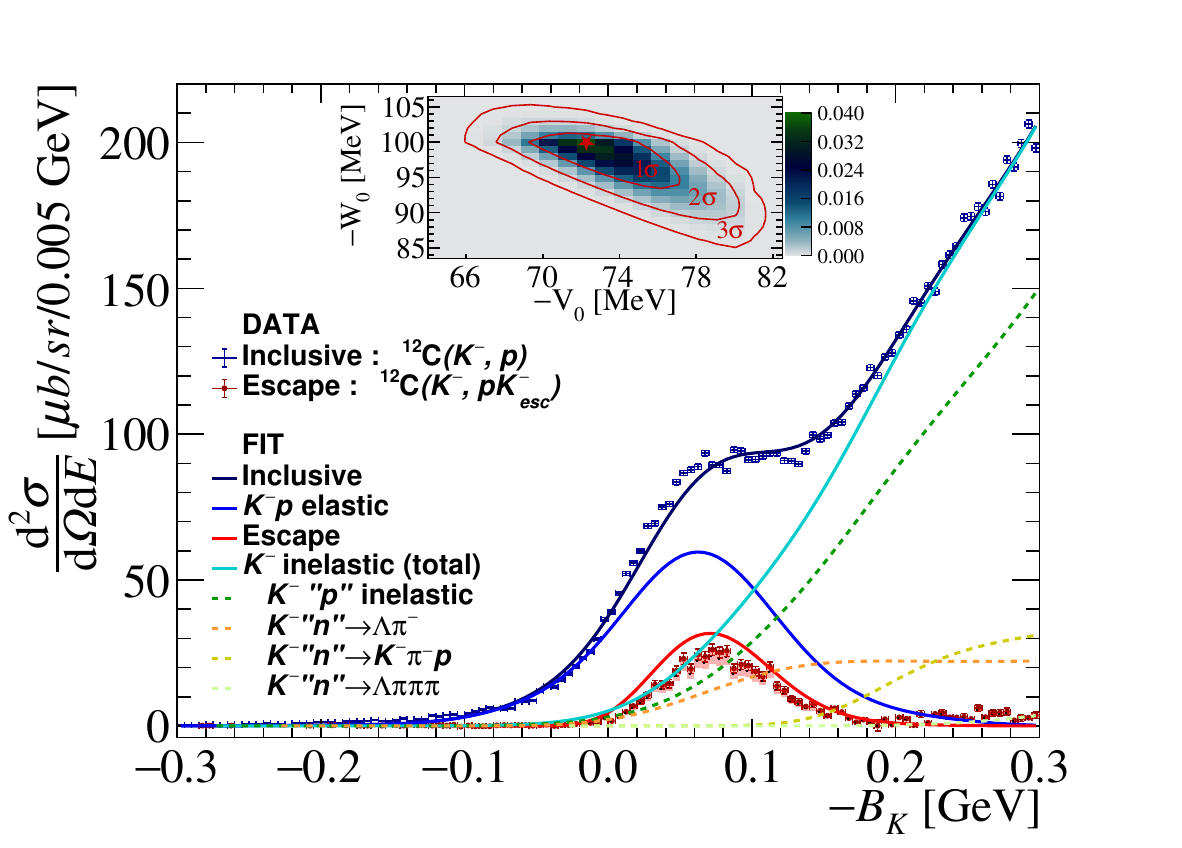}}
\caption{
Double differential cross sections of $^{12}{\rm C}(K^-,p)$ at 1.8 GeV/$c$ vs.\ $\bar{K}$ binding energy $-B_K$. Blue (red) points show the inclusive (escape) data. The solid curves show the total inclusive fit, its quasi-free $K^-``p"$ elastic and total $K^-$ inelastic components, and the escape fit; the dashed curves show the individual inelastic contributions, as labeled in the legend. The pink intervals associated with the escape data points indicate their systematic uncertainties. The inset shows the posterior probability density in the $(-V_0,-W_0)$ plane (color scale, arbitrary units) with the $1\sigma$, $2\sigma$, and $3\sigma$ HPD contours; the star indicates the best-fit point.
}
\label{fig_spectrum}
\end{figure}

\textbf{\textit{Results and Discussion.---}} Figure~\ref{fig_spectrum} shows the simultaneously measured inclusive (blue) and $K^-$-escape (red) $^{12}{\rm C}(K^-,p)$ missing-mass spectra as double differential cross sections.

The curves in Fig.~\ref{fig_spectrum} represent the best-fit results. The inclusive data are described by the total fit, the sum of the quasi-free $K^-``p"$ elastic component from the Green's function calculation and the total $K^-$ inelastic background; the latter is decomposed into the quasi-free $K^-``p"$ inelastic contribution and the $K^-``n"$ channels ($\to\Lambda\pi^-$, $\to K^-\pi^-p$, and $\to\Lambda\pi\pi\pi$), shown as dashed curves and labeled in the legend. For the escape data, the best-fit curve describes the theoretical $K^-$ escape spectrum; the pink intervals associated with the escape data points indicate their systematic uncertainties. The simultaneous fit removes the large uncertainty in $W_0$ that persisted in the previous inclusive measurement. A localized minimum is obtained in the likelihood map shown in the inset, with $1\sigma$, $2\sigma$, and $3\sigma$ highest-posterior-density (HPD) contours, yielding $$V_0=-72\:^{+3}_{-5}\:(\text{stat.})\:^{+0}_{-8}\:(\text{syst.})~\rm{MeV},$$$$W_0=-100\:^{+7}_{-1}\:(\text{stat.})\:^{+0}_{-16}\:(\text{syst.})~\rm{MeV}.$$ As for uncertainties due to the theoretical model, varying the distortion cross section used in the eikonal approximation ($\pm10\%$) and the density dependence of the optical potential following Mare\v{s} \textit{et al.}~\cite{Mares2006}, we find that only the distortion variation shifts $V_0$ (by at most $\pm7$~MeV), while $W_0$ remains unchanged.

The systematic uncertainties on $V_0$ and $W_0$ were evaluated by scaling the escape-spectrum normalization within its total uncertainty of $\pm10.2\%$ and repeating the simultaneous fit. Among these conditions, the one yielding the largest likelihood was adopted as the central value, and the maximum shift in each direction was quoted as the systematic error. All other conditions shifted the best-fit values only toward more negative $V_0$ and $W_0$, resulting in the one-sided uncertainties $^{+0}_{-8}$ and $^{+0}_{-16}$~MeV. Stability was further confirmed by varying the relative fractions of the individual $K^-$ inelastic channels and the fit range around the nominal $-0.04\le-B_K\le0.2$~GeV. In all cases, $V_0$ and $W_0$ shifted by less than 1~MeV---negligible compared with the quoted systematic uncertainties---while only the inelastic-background composition was redistributed.

Converted to normal nuclear density $\rho_0$, the present result corresponds to $(-67, -93)$ MeV, while the J-PARC E05 result~\cite{Ichikawa2020} gives $\sim (-74, -37)$ MeV. The real parts are mutually consistent. On the other hand, the present experiment with the exclusive escape measurement yields a substantially stronger imaginary part than J-PARC E05, which relies solely on the inclusive spectrum and thus cannot directly constrain $W_0$.

It is important to compare our result with those derived from kaonic atom X-ray spectroscopy. A recent global analysis~\cite{Yamagata-Sekihara2024} incorporating the latest precision data~\cite{Hashimoto2022} alongside heavier kaonic atom data favors a shallow absorptive solution, $(-90,-120)$~MeV. Here, kaonic atom spectroscopy probes the interaction in the energy region near $B_K \approx 0$, whereas the present in-flight measurement probes a broader range of $-B_K = 0$--$0.2$~GeV. Although core-nucleus breakup reactions induced by $K^-``N" \to \bar{K}N$ rescattering additionally contribute to $W_0$ at $-B_K > 0$, both results consistently point to a strongly absorptive potential ($W_0 \lesssim -100$~MeV) accompanied by a shallow real part. The present volume-sensitive measurement thus resolves part of the long-standing ambiguity---the deep-real solutions are disfavored---while the energy dependence of $W_0$ and the microscopic decomposition of the absorptive strength remain to be clarified.

According to these comparisons, the shallow real part ($V_0=-72$~MeV) is close to the chiral-unitary prediction~\cite{Kaiser1995,Waas1997} ($V_0\sim-50$ to $-80$~MeV), while the imaginary part ($W_0=-100$~MeV) substantially exceeds its prediction ($W_0\sim-40$~MeV) based on single-nucleon absorption ($K^-N\to Y\pi$). This is consistent with significant contributions beyond one-nucleon absorption, such as multi-nucleon absorption and $K^-``N"\to\bar{K}N$ rescattering/core-breakup channels already invoked above; a unique identification of the dominant mechanism is beyond the scope of this Letter.

The large absorptive strength is also qualitatively consistent with the $K^-pp$ system observed by J-PARC E15~\cite{Ajimura2019,Hashimoto2015,Yamaga2020}, where multi-nucleon absorption channels have been discussed~\cite{Sekihara:2016,Dote:2018}.

These parameters may affect kaon condensation in neutron stars, but such implications remain indirect and model dependent because no NS calculations are presented here~\cite{Hong2024,Muto2021,GuhaRoy2025}. They qualitatively favor a higher onset density for $K^-$ condensation than deep-potential scenarios.

In conclusion, by performing the first exclusive measurement of the $K^-$ escape process and achieving a simultaneous fit with the inclusive spectrum, we have provided the first direct experimental constraint on the imaginary potential strength $W_0$. The present result will play a key role in understanding antikaon absorption dynamics in nuclear matter.

\begin{acknowledgments}
The authors would like to thank the J-PARC accelerator staff and the Hadron Experimental Facility for their excellent support in providing a stable $K^-$ beam and maintaining the experimental conditions. We also thank T.~Muto for valuable discussions on the implications for neutron star physics. This work was supported by JSPS KAKENHI Grant Number JP23K20852 (Grant-in-Aid for Scientific Research (B)), 21H00130, and 21H04478, by the Ministry of Education, Culture, Sports, Science and Technology (MEXT) of Japan Grant-in-Aid for Innovative Areas (18H05403), and by the National Research Foundation of Korea (NRF) grant RS-2024-00355188. The present work was conducted under the Reimei Research Program of the Japan Atomic Energy Agency. One of the authors (F.~Oura) was supported by JSPS Research Fellowship for Young Scientists (Grant Number JP24KJ0398).
\end{acknowledgments}

\bibliography{ref}

@article{Jido2003,
  author    = {D.~Jido and J.~A.~Oller and E.~Oset and A.~Ramos and U.~-G.~Meissner},
  title     = {Chiral dynamics of the two {$\Lambda(1405)$} states},
  journal   = {Nucl. Phys. A},
  volume    = {725},
  pages     = {181--200},
  year      = {2003},
  doi       = {10.1016/S0375-9474(03)01598-7}
}

@article{Martin1981,
  author  = {A.~D.~Martin},
  title   = {Kaon-nucleon parameters},
  journal = {Nuclear Physics B},
  volume  = {179},
  pages   = {33--48},
  year    = {1981},
  doi     = {10.1016/0550-3213(81)90247-9}
}

@article{IkedaHyodoWeise2012,
  author  = {Y.~Ikeda and T.~Hyodo and W.~Weise},
  title   = {Chiral {SU}(3) theory of antikaon-nucleon interactions with improved threshold constraints},
  journal = {Nuclear Physics A},
  volume  = {881},
  pages   = {98--114},
  year    = {2012},
  doi     = {10.1016/j.nuclphysa.2012.01.012}
}

@article{Hyodo2013,
  author  = {T.~Hyodo},
  title   = {{A}ntikaon-nucleon dynamics and its application to few-body systems},
  journal = {Nuclear Physics A},
  volume  = {914},
  pages   = {260--274},
  year    = {2013},
  doi     = {10.1016/j.nuclphysa.2013.01.015}
}

@article{Bazzi2011,
  author  = {M.~Bazzi and others},
  title   = {A new measurement of kaonic hydrogen {X}-rays},
  journal = {Physics Letters B},
  volume  = {704},
  pages   = {113--117},
  year    = {2011},
  doi     = {10.1016/j.physletb.2011.09.011}
}

@article{Iwasaki1997,
  title = {Observation of Kaonic Hydrogen {${K}_{\ensuremath{\alpha}}$ X Rays}},
  author = {Iwasaki, M. and others},
  journal = {Phys. Rev. Lett.},
  volume = {78},
  issue = {16},
  pages = {3067--3069},
  numpages = {0},
  year = {1997},
  month = {Apr},
  publisher = {American Physical Society},
  doi = {10.1103/PhysRevLett.78.3067},
  url = {https://link.aps.org/doi/10.1103/PhysRevLett.78.3067}
}

@article{AkaishiYamazaki2002,
  author  = {Y.~Akaishi and T.~Yamazaki},
  title   = {{N}uclear {$\bar{K}$} bound states in light nuclei},
  journal = {Physical Review C},
  volume  = {65},
  number  = {4},
  pages   = {044005},
  year    = {2002},
  doi     = {10.1103/PhysRevC.65.044005}
}

@article{Ajimura2019,
    author = "Ajimura, S. and others",
    collaboration = "J-PARC E15",
    title = "{``$K^-pp$'', a $\bar{K}$-meson nuclear bound state, observed in $^3$He($K^-$, $\Lambda p$)$n$ reactions}",
    eprint = "1805.12275",
    archivePrefix = "arXiv",
    primaryClass = "nucl-ex",
    doi = "10.1016/j.physletb.2018.12.058",
    journal = "Phys. Lett. B",
    volume = "789",
    pages = "620-625",
    year = "2019"
}

@article{Hashimoto2015,
    author = "Hashimoto, T. and others",
    collaboration = "J-PARC E15",
    title = "{Search for the deeply bound $K^-pp$ state from the semi-inclusive forward-neutron spectrum in the in-flight $K^-$ reaction on helium-3}",
    doi = "10.1093/ptep/ptv076",
    journal = "Prog. Theor. Exp. Phys.",
    volume = "2015",
    number = "6",
    pages = "061D01",
    year = "2015"
}

@article{Yamaga2020,
    author = "Yamaga, T. and others",
    collaboration = "J-PARC E15",
    title = "{Observation of a $\bar{K}NN$ bound state in the $^3\mathrm{He}(K^-,n)$ reaction}",
    doi = "10.1103/PhysRevC.102.044002",
    journal = "Phys. Rev. C",
    volume = "102",
    number = "4",
    pages = "044002",
    year = "2020"
}

@article{Sekihara:2016,
  author  = {Takayasu Sekihara and Eulogio Oset and Angel Ramos},
  title   = {{On the structure of the $K^-pp$ resonance in the $^3$He($K^-, \Lambda p$)n reaction}},
  journal = {Prog. Theor. Exp. Phys.},
  volume  = {2016},
  pages   = {123D03},
  year    = {2016},
  doi     = {10.1093/ptep/ptw166}
}

@article{Dote:2018,
  author  = {Akinobu Dot{\'e} and Takashi Inoue and Takayuki Myo},
  title   = {{Investigation of the $K^-pp$ bound state with a coupled-channel complex scaling method}},
  journal = {Phys. Lett. B},
  volume  = {784},
  pages   = {197--202},
  year    = {2018},
  doi     = {10.1016/j.physletb.2018.07.056}
}

@article{Batty1997,
  author = {C.J. Batty and E. Friedman and A. Gal},
  title = {Strong interaction physics from hadronic atoms},
  journal = {Physics Reports},
  volume = {287},
  number = {5},
  pages = {385-445},
  year = {1997},
  issn = {0370-1573},
  doi = {https://doi.org/10.1016/S0370-1573(97)00011-2},
  url = {https://www.sciencedirect.com/science/article/pii/S0370157397000112}
}

@article{Mares2006,
  author    = {J.~Mare\v{s} and E.~Friedman and A.~Gal},
  title     = {{$\bar{K}$}-nuclear bound states in a dynamical model},
  journal   = {Nucl. Phys. A},
  volume    = {770},
  pages     = {84--105},
  year      = {2006},
  doi       = {10.1016/j.nuclphysa.2006.02.010}
}

@article{Friedman2007,
  author    = {E.~Friedman and A.~Gal},
  title     = {In-medium nuclear interactions of low-energy hadrons},
  journal   = {Physics Reports},
  volume    = {452},
  pages     = {89--153},
  year      = {2007},
  doi       = {10.1016/j.physrep.2007.04.001}
}

@article{Kishimoto2007,
  author  = {T.~Kishimoto and others},
  title   = {Search for deeply bound kaonic nuclear states by in-flight $K^-$ reactions},
  journal = {Prog. Theor. Phys.},
  volume  = {118},
  pages   = {181--202},
  year    = {2007},
  doi     = {10.1143/PTP.118.181}
}

@article{Ichikawa2020,
  author = {Y.~Ichikawa and others}, 
  collaboration = {J-PARC E05 collaboration},    
  title = {An event excess observed in the deeply bound region of the {$^{12}{\rm C}(K^-, p)$} missing-mass spectrum},
  journal = {Progress of Theoretical and Experimental Physics},
  volume = {2020},
  number = {12},
  pages = {123D01},
  year = {2020},
  doi = {10.1093/ptep/ptaa139}
}

@article{Jung2025,
  author = {W.~S.~Jung and others},
  collaboration = {J-PARC E42 Collaboration},
  title = {{C}ross {S}ection {M}easurements for {$^{12}{\rm C}(K^-, K^+\Xi^-)$} and {$^{12}{\rm C}(K^-, K^+\Lambda\Lambda)$} {R}eactions at 1.8 {GeV}/{$c$}},
  journal = {Progress of Theoretical and Experimental Physics},
  volume = {2025},
  number = {9},
  pages = {091D01},
  year = {2025},
  doi = {10.1093/ptep/ptaf064}
}

@article{Antonov2002,
  author  = {A.~N.~Antonov and M.~K.~Gaidarov and M.~V.~Ivanov and D.~N.~Kadrev and G.~Z.~Krumova and P.~E.~Hodgson and H.~V.~von~Geramb},
  title   = {Nucleon momentum distribution in deuteron and other nuclei within the light-front dynamics method},
  journal = {Phys. Rev. C},
  volume  = {65},
  pages   = {024306},
  year    = {2002},
  doi     = {10.1103/PhysRevC.65.024306}
}

@article{Agostinelli2003,
  author  = {S.~Agostinelli and others},
  title   = {{Geant4}---a simulation toolkit},
  journal = {Nucl. Instrum. Methods Phys. Res. A},
  volume  = {506},
  pages   = {250--303},
  year    = {2003},
  doi     = {10.1016/S0168-9002(03)01368-8}
}

@article{Yamagata2006,
  title = {In-flight (${K}^{\ensuremath{-}},p$) reactions for the formation of kaonic atoms and kaonic nuclei using the Green function method},
  author = {Yamagata, J. and Nagahiro, H. and Hirenzaki, S.},
  journal = {Phys. Rev. C},
  volume = {74},
  issue = {1},
  pages = {014604},
  numpages = {9},
  year = {2006},
  month = {Jul},
  publisher = {American Physical Society},
  doi = {10.1103/PhysRevC.74.014604},
  url = {https://link.aps.org/doi/10.1103/PhysRevC.74.014604}
}

@article{Kaiser1995,
  author    = {N.~Kaiser and P.~B.~Siegel and W.~Weise},
  title     = {Chiral dynamics and the low-energy kaon-nucleon interaction},
  journal   = {Nuclear Physics A},
  volume    = {594},
  pages     = {325--345},
  year      = {1995},
  doi       = {10.1016/0375-9474(95)00362-5}
}

@article{Waas1997,
  author    = {T.~Waas and W.~Weise},
  title     = {{S}-wave interactions of {$\bar{K}$} and {$\eta$} mesons in nuclear matter},
  journal   = {Nuclear Physics A},
  volume    = {625},
  pages     = {287--306},
  year      = {1997},
  doi       = {10.1016/S0375-9474(97)00487-9}
}

@article{Hashimoto2022,
  author = {T.~Hashimoto and others},
  collaboration = {J-PARC E62 Collaboration},
  title = {{M}easurements of {S}trong-{I}nteraction {E}ffects in {K}aonic-{H}elium {I}sotopes at {S}ub-e{V} {P}recision with {X}-{R}ay {M}icrocalorimeters},
  journal = {Phys. Rev. Lett.},
  volume = {128},
  issue = {11},
  pages = {112503},
  year = {2022},
  doi = {10.1103/PhysRevLett.128.112503}
}

@article{Yamagata-Sekihara2024,
    author = {J.~Yamagata-Sekihara and Y.~Iizawa and D.~Jido and N.~Ikeno and T.~Hashimoto and S.~Okada and S.~Hirenzaki},
    title = {{I}nvestigation of {K}aonic {A}tom {O}ptical {P}otential by the {H}igh-{P}recision {D}ata of {K}aonic {$^3$He} and {$^4$He} {A}toms},
    journal = {Progress of Theoretical and Experimental Physics},
    volume = {2025},
    number = {1},
    pages = {013D02},
    year = {2024},
    month = {12},
    doi = {10.1093/ptep/ptae189},
}

@article{Muto2021,
  author  = {T.~Muto and T.~Maruyama and T.~Tatsumi},
  title   = {{E}ffects of three-baryon forces on kaon condensation in hyperon-mixed matter},
  journal = {Phys. Lett. B},
  volume  = {821},
  pages   = {136587},
  year    = {2021},
  doi     = {10.1016/j.physletb.2021.136587}
}

@article{VeldeWilquet1977,
  author  = {C.~Vander Velde-Wilquet and J.~Sacton and J.~H.~Wickens and D.~N.~Tovee and D.~H.~Davis},
  title   = {{D}etermination of the branching fractions for {$K^-$} meson absorption at rest in carbon nuclei},
  journal = {Nuovo Cimento A},
  volume  = {39},
  number  = {4},
  pages   = {538--547},
  year    = {1977},
  doi     = {10.1007/BF02804939}
}

@article{Hong2024,
  author = {Bin Hong and Zhongzhou Ren},
  title = {Probing kaon meson condensations through gravitational waves during neutron star inspiral phases},
  journal = {Physics Letters B},
  volume = {858},
  pages = {139076},
  year = {2024},
  issn = {0370-2693},
  doi = {https://doi.org/10.1016/j.physletb.2024.139076},
  url = {https://www.sciencedirect.com/science/article/pii/S0370269324006348},
}

@article{GuhaRoy2025,
  author = {Guha Roy, Debanjan and Banik, Sarmistha},
  title = {{S}ignatures of {$K^-$} condensation on neutron star structure and {$f$}-mode frequencies},
  journal = {Physics Letters B},
  volume = {871},
  pages = {139990},
  year = {2025},
  issn = {0370-2693},
  doi = {https://doi.org/10.1016/j.physletb.2025.139990},
  url = {https://www.sciencedirect.com/science/article/pii/S0370269325007488}
}

@article{Beer2005,
  author = {Beer, G. and others},
  collaboration = {DEAR Collaboration},
  title = {{M}easurement of the {K}aonic {H}ydrogen {X}-{R}ay {S}pectrum},
  journal = {Phys. Rev. Lett.},
  volume = {94},
  pages = {212302},
  year = {2005},
  doi = {10.1103/PhysRevLett.94.212302}
}

@article{Takahashi2012,
  author  = {T.~Takahashi and others},
  title   = {Beam and SKS spectrometers at the K1.8 beam line},
  journal = {Progress of Theoretical and Experimental Physics},
  volume  = {2012},
  pages   = {02B010},
  year    = {2012},
  doi     = {10.1093/ptep/pts023},
  url     = {https://doi.org/10.1093/ptep/pts023}
}

@article{Ahn_2023, title={Superconducting dipole magnet for Hyperon spectrometer}, volume={1047}, ISSN={0168-9002}, url={http://dx.doi.org/10.1016/j.nima.2022.167775}, DOI={10.1016/j.nima.2022.167775}, journal={Nuclear Instruments and Methods in Physics Research Section A: Accelerators, Spectrometers, Detectors and Associated Equipment}, publisher={Elsevier BV}, author={Ahn, J.K. and Choi, S. and Hasegawa, S. and Hayakawa, S.H. and Hong, J. and Ichikawa, Y. and Imai, K. and Kim, S.H. and Makida, Y. and Ohhata, H. and Sako, H. and Sasaki, K. and Sato, S. and Takahashi, T. and Tanida, K. and Yoshida, J.}, year={2023}, month=feb, pages={167775} }

@article{Kim_2019, title={High-rate performance of a time projection chamber for an H-dibaryon search experiment at J-PARC}, volume={940}, ISSN={0168-9002}, url={http://dx.doi.org/10.1016/j.nima.2019.06.050}, DOI={10.1016/j.nima.2019.06.050}, journal={Nuclear Instruments and Methods in Physics Research Section A: Accelerators, Spectrometers, Detectors and Associated Equipment}, publisher={Elsevier BV}, author={Kim, S.H. and Ichikawa, Y. and Sako, H. and Ahn, J.K. and Akaishi, T. and Ashikaga, S. and Choi, S.W. and Ekawa, H. and Hasegawa, S. and Hayakawa, S. and Jung, W.S. and Kang, B.M. and Lee, J.Y. and Nanamura, T. and Sato, S. and Shirotori, K. and Suzuki, K. and Tanida, K. and Yang, S.B. and Yoshida, J.}, year={2019}, month=oct, pages={359–370} }

@article{Agari_2012, title={Secondary charged beam lines at the J-PARC hadron experimental hall}, volume={2012}, ISSN={2050-3911}, url={http://dx.doi.org/10.1093/ptep/pts038}, DOI={10.1093/ptep/pts038}, number={1}, journal={Progress of Theoretical and Experimental Physics}, publisher={Oxford University Press (OUP)}, author={K.~Agari and others}, year={2012} }

\end{document}